Atomic parity violation and the standard model

Carl Wieman
Stanford University, Stanford, CA, 94305

Andrei Derevianko
University of Nevada, Reno, NV 89557

**Introduction**

Parity violation in an atomic system can be observed as an electric dipole transition amplitude between two atomic states with the same parity, such as the $6S$ and $7S$ states in cesium. This transition amplitude can be written as $E_{PV} = k\, Q_W$, where $k$ is an atomic-structure factor, and $Q_W$ is the weak charge, which can be expressed in terms of standard model coupling constants. The experimental measurement of $E_{PV}$ and the calculation of $k$ are equivalently important for the precise determination of $Q_W$ as a test of the standard model. While parity violation has been measured in a variety of different atoms, for the past three-plus decades, the precision of the combination of the experimental measurement of $E_{PV}$ and the theoretical calculation of $k$ has been the best for the case of atomic cesium. One of us (CW) has focused on the experimental measurements, while the other (AD) has advanced the theoretical calculation of $k$. For the past 20 years, the fractional uncertainty in the experimental measurement in cesium has remained at 0.35 percent level, and the atomic theory has been slowly and steadily improving, so it is now comparable.

At various times in the past, depending on the status of the cesium work and collider experiments, the value of $Q_W$ from PV in cesium has provided the most stringent limits on various possible types of "new" physics. Atomic PV is uniquely sensitive to extra $Z$ bosons, such as those predicted in grand unified theories, technicolor models, SUSY, and string theories. Generally, atomic PV is one of a few probes of electroweak coupling below the $Z-$pole and one of the few sensitive to electron-neutron couplings. While the relevant momentum transfer is just 30 MeV, the exquisite accuracy of the interpretation constrains new physics at much higher energies. The present determination of the $Q_W$ ($^{133}$Cs) probes the new physics at a mass scale of $\sim 20$ TeV.

The interest in using atomic parity violation to explore the standard model began with the 1974 papers by the Bouchiats [1–3] which pointed out the potential for looking for weak neutral currents predicted by standard model by looking at their effects on parity violation in atoms[1]. They showed that the short-range nature of interaction would result in $Z^3$ enhancement, where $Z$ in the atomic number. This enhancement factor made what seemed to be an impossibly small effect into something that might conceivably be detectable (albeit with great difficulty), as electric dipole transitions between atomic states with same parity for atoms with large $Z$. The Bouchiats also proposed general experimental approaches for how these might be observed, particularly the interference with electric field induced weak E1 transitions, such as in cesium.

This led to a flurry of activity. During the late 70's and early 80s there were a number of experiments [4–6] looking at optical rotation of light in atomic vapor cells. The frequency of the light was scanned across allowed M1 transitions, notably in bismuth [4]. These experiments produced a variety of conflicting results, as did the attempts to calculate the atomic structure factor for the various transitions in bismuth, a rather complex atom.  This gave little confidence that atomic parity violation would provide a useful test of the standard model.  With the measurement of PV in deep inelastic scattering by Prescott *et al* [7], the basic observation of atomic PV as confirmation of the existence of weak neutral currents in the standard model became less relevant.

---

[1] Zel'dovich [30] contemplated a possibility of observing PV in atoms in 1959. He concluded that the effect was too small to be of experimental significance.

Over the next several years however, there were a number of measurements of atomic PV in multiple atoms, either by optical rotation experiments with better understanding of the systematic errors, or measurements of PV in the excitation rates of electric field induced E1 transitions.  These had uncertainties, both in experiments and/or calculations of $k$, at the 10-20% level.  Several reviews have been written on this early record of atomic PV and a broader discussion of atomic PV (see Refs. [8–12]). These usually have somewhat sanitized versions of the numbers actually produced, leaving out the degree of disagreement and uncertainty of the early results at the time, which left most elementary particle physicists appropriately skeptical as to the usefulness of such results for testing the standard model.

At that time, one of us (CW) was just starting to work on atomic PV. At that time, it was clear that if a measurement of atomic PV was to be meaningful in terms of standard model physics, it would require a different focus than earlier work, where the goal was simply to observe PV in atoms.  It would require high precision measurements in an atom where the atomic theory could also be done and tested with high precision.  Another important shift in focus from much of the earlier work was that potential systematic errors, which had resulted in much of the variation and inconsistency in the early work, were at least as important in the experimental design as the signal to noise ratio.  This resulted in the design choices of cesium as the atom for measurement, and specifically the $6S$ to $7S$ transition, and a relatively elaborate experimental design involving an atomic beam and lots of elaborate laser technology.  Those design considerations proved to be appropriate, as the experimental measurements ultimately proved considerably more precise than any others.  The corresponding atomic theory in cesium has now similarly been carried forward to unprecedented precision.  This progress required major advances in experimental and theoretical methods.  The result has been testing of the standard model physics at the parts in $10^3$ level, at a different energy scale than collider experiments and with different sensitivity to new (i.e. beyond standard model) physics.  As this volume is devoted to the standard model, we will confine the rest of our discussion to this cesium work.

**Experimental design**
Although the details are complicated, the basic concepts behind PV in atoms, and its measurement are quite simple.  The neutral current interaction arising from the exchange of a $Z$ boson between the electrons and the nucleons in an atom leads to a mixing of the $S$ and $P$ states in the atoms.  The amount of mixing scales roughly as $Z^3$, but even for large $Z$ it is only about 1 part in $10^{11}$ .  This mixing is detected by observing the electric dipole transition amplitude between two $S$ (or two $P$) states in an atom.  Such an amplitude can only exist if parity is violated.  Because the amplitude is so small, it can only be observed by using interference techniques.  In such a technique the $nS$ to $n'S$ (or $P$) transition rate is given by

$$R = |A_0 \pm A_{pv}|^2 = A_0^2 \pm \mathbf{2A_0 A_{pv}} + A_{pv}^2,$$

where $A_{pv}$ is the desired PV amplitude and $A_0$ is a larger parity conserving amplitude.  Because the interference term is linear in $A_{pv}$ it can be large enough to measure, however it must be distinguished from the large background due to the $A_0^2$ contribution.  For there to be a nonzero interference term, the experiment must have a "handedness", and if the handedness is reversed, the interference term will change sign, and can thereby be distinguished as a modulation in the transition rate.

4The 1997 measurement [13] of PV in cesium was the culmination of 14 years of work, which had produced a series of increasingly precise measurements. The Stark interference technique is used to detect the PV E1 amplitude on the $6S$ to $7S$ transition. This transition is excited in an atomic beam by a laser that intersects the beam at right angles. The directions of three perpendicular vectors, that of an applied electric field, an applied DC magnetic field, and the angular momentum of the laser photon, define the handedness of the apparatus. The PV interference term then changes sign, causing a modulation in the excitation rate, as this coordinate system is reversed back and forth between right and left handed. There are four such reversals: changing sign of each of the E and B fields, reversing the laser light from left to right circular polarization, and reversing the sign of the m level that is being excited. This use of multiple redundant reversals greatly suppresses possible systematic errors.

In the experiment, an intense beam of atomic cesium passes through beams from two diode lasers that optically pump it into a single $F, M_F$ state. It then continues on to intersect an intense green standing wave laser field between electric field plates. The standing wave is produced by a Fabry-Perot power buildup cavity that increases the incident laser power by a factor of 30,000 when the laser frequency is locked in resonance with the cavity. After the interaction region, the beam goes into the detection region where it intersects another diode laser beam that drives a cycling transition for the atoms in the previously emptied $F$ level. The resulting florescence is detected on a photodiode. There are a number of aspects of this apparatus that are rather challenging. First it requires five lasers (four diode, one dye) that must be stabilized to nearly state of the art performance (both for 25 years ago and still today). This means frequency stabilities on the order of 1 part in $10^{14}$ /s$^{1/2}$ and intensity stabilities of 1 part in $10^6$/s$^{1/2}$. The power buildup cavity also involves some challenges, since the spacing between the mirrors must be held constant to about 1/100 of an atomic diameter per s$^{1/2}$. Meeting any one of these requirements was not easy, but by far the most difficult aspect of the technology is that all these things must work at the same time, and they must do that for hours, days, and months in order to acquire the necessary data.

The bulk of the experimental running time, some 7000 hours, and effort spent on this experiment was devoted to the study and elimination of potential systematic errors. This involved taking data under a wide variety of conditions and looking at the signals with all the possible combinations of different field reversals. A number of statistical analyses of the data were also carried out. The resulting systematic errors are all below 0.1% of the PV signal.

The final result for the parity violating amplitude on the two different hyperfine lines measured is:

$$\text{Im}\left(\frac{E_{PV}}{\beta}\right) = \begin{cases} 1.6349(80) \text{ mV/cm}, & F = 4 \text{ to } F' = 3 \\ 1.5576(77) \text{ mV/cm}, & F = 3 \text{ to } F' = 4 \end{cases}.$$

The PV amplitude is in units of the equivalent electric field required to give the same mixing of $S$ and $P$ states as the PV interaction. Below we show a summary of all the measurements of PV in cesium.

Table 1 **Measurements of parity violating amplitude for the $6S - 7S$ transition in cesium**

| Experiment | $\text{Im}(E_{pv}/\beta)$, mV/cm | |
|---|---|---|
| Paris 1982 [14] | 1.34 ± 0.22 ± 0.11 | |
| Paris '86 [15] | 1.52 ± 0.18 | |
| Colorado '86 [16] | 1.65 ± 0.13 | |
| Colorado '88 [17] | 1.639 ± 0.047 | $F = 4$ to $F' = 3$ |
| | 1.513 ± 0.049 | $F = 3$ to $F' = 4$ |
| | 1.576 ± 0.034 | combined |
| Colorado '97 [13] | 1.6349 ± 0.0080 | $F = 4$ to $F' = 3$ |
| | 1.5576 ± 0.0077 | $F = 3$ to $F' = 4$ |
| | 1.5935 ± 0.0056 | combined |

It can be seen that all the experiments are consistent, and that the first hint of a hyperfine dependence seen in the Colorado 1988 result [17] was confirmed by the later work [13]. The measured difference is 0.077 (11) mV/cm. The primary difference between these two hyperfine lines is that the nuclear spin is reversed relative to the electron spin. This observed difference is a measure of the nuclear spin dependent contribution to atomic parity violation, and as such is the first, and still only, observation of an "anapole moment" [18,19].

From the standard model perspective, the main significance is that atomic PV is measured on two transitions corresponding to a nearly exact reversal of the anapole contribution. This allows us to cancel out the poorly understood nuclear dependent parts by taking an appropriately weighted average of the two lines. The result is a quantity, 1.5935(56) mV/cm = $k\, Q_W$, that is directly related to standard model coupling constants through $Q_w$ and the atomic structure factor, $k$.

**Atomic theory for the structure factor**
When the above experimental result was first obtained, its value as a test of the standard model was limited by the uncertainty with which the atomic structure factor $k$ could be calculated. Since that time, advances in the atomic theory have led to improved calculations. That has resulted in theoretical uncertainties in $k$ that are now similar to those of the experiment. This leads to an inferred nuclear weak charge [20], $Q_W(^{133}Cs) = -72.58(29)_{expt}(32)_{theor}$. This agrees with the prediction of the standard model at the $1\sigma$ level.

The PV amplitude of the $6S - 7S$ transition can be expressed as

$$E_{PV} = \sum_n \frac{\langle 7S|D|nP\rangle\langle nP|H_W|6S\rangle}{E_{6S} - E_{nP}} + (7S \leftrightarrow 6S), \quad (1)$$

where $D$ and $H_W$ are dipole and weak interaction operators. This sum is evaluated numerically by solving the atomic-structure problem. For the accurate summation, the approximate wave-functions have to reproduce the short-range (close to the nucleus) and long-range atomic properties simultaneously. High accuracy evaluation of this sum and the supporting properties of Cs atom have helped drive the development of *ab initio* relativistic methods in atomic theory. These improved methods have found applications beyond atomic parity violation.

An alkali-metal atom, Cs has a single valence electron outside a tightly bound Xe-like core. The relative hydrogen-like simplicity of this atom is crucial for reaching the desired ~0.1% theoretical accuracy. Such a many-electron problem as the full cesium atom is very difficult; in classical mechanics even the much simpler three-body problem cannot be solved in closed form. For the Cs atom, one needs to solve for a correlated motion of 55 electrons. Even storing the resulting many-body wave function on a sparse 10-point grid requires $10^{55\times 3}$ points, which greatly exceeds the estimated number of atoms in the observable Universe. Moreover, the matrix elements of the weak interaction are accumulated inside the nucleus, where the electrons are relativistic, $v/c \sim \alpha Z \sim 0.4$. Therefore, the treatment must be *ab initio* relativistic.

The most successful approaches are rooted in the applications of relativistic many-body perturbation theory (MBPT). These approaches start with a mean-field Dirac-Hartree-Fock picture of an atom and treat the residual $e^- - e^-$ Coulomb interaction as a perturbation. Application of MBPT is essential, as illustrated by the fact that the lowest-order value for the hyperfine-structure constant of the $6S$ state is $\sim 50\%$ off from the experimental value. This constant describes the strength of coupling of the electrons to the nuclear magnetic moment and its short-distance nature mimics the behavior of the weak matrix elements. In the most successful applications of MBPT, major classes of topologically similar MBPT diagrams are summed to all orders of perturbation theory. The dominant effects come from the valence electron polarizing the atomic core and from the external-field induced rearrangement of the electronic cloud.

The calculations are precisely tested by comparing the calculated values with measured values for energies, hyperfine structure constants, and electric-dipole matrix elements. This is another way in which cesium is a particularly desirable atom to use for PV studies, as many of its atomic properties have been measured very precisely, much better than any other heavy atom, likely because of its role in defining the second.

The successful development of relativistic MBPT, both conceptually and numerically, lead to an important 1% accuracy milestone in determination of the structure factor $k$. This milestone was reached by the Novosibirsk [21] and Notre Dame [22] groups in the late 1980s. Both groups have pursued the so called "parity-mixed" approach, where the weak interaction is combined with the atomic Hamiltonian, and the MBPT calculations are carried out using the resulting parity-mixed basis of atomic orbitals. In addition to the parity-mixed approach, the Notre Dame group has employed the coupled cluster method, which offers a systematic way of summing diagrams to all orders using iterative techniques.

New atomic lifetime and polarizability data reported in 1999 improved the theory-experiment agreement for these values, and based on this improved agreement, the theoretical uncertainty in the structure factor $k$ was reduced to 0.4% [23]. The resulting value of the Cs weak charge differed by 2.5 $\sigma$ from the standard model prediction. The reduced theoretical uncertainty raised the question of whether some small sub-1% atomic-structure effects could be the reason for the discrepancy. Several groups contributed to understanding such small corrections, see review [24] and references therein. The dominant corrections were found to be due to the Breit interaction, radiative QED effects, and the neutron skin correction, which is the difference between the well-known proton nuclear distribution and the relatively poorly known neutron distribution that enters into $H_W$. In 2005, these corrections essentially reconciled parity violation in Cs with the standard model, with theoretical uncertainty standing at 0.5%, still larger than the experimental error bar.

In parallel with the evaluation of ``small'' corrections, the efforts of the AD group have focused on the more accurate solution of the many-body problem. This has been done in the systematic framework of coupled-cluster formalism, extending the earlier efforts by the Notre Dame group [22]. The Notre Dame calculations were complete through the third order of MBPT. The next step was to design a numerically-tractable formalism that would recover all fourth-order diagrams and, in addition, sum certain classes to all orders. The challenge is that while the computational complexity increases, the intuition in selecting many important diagrams at the 0.1% level may fail (usually, it is easier to justify the selection after the result is known). To motivate the next-generation formalism, AD and collaborators have explicitly computed 1648 fourth-order diagrams for matrix elements [25]. That exercise and the developed toolbox guided and enabled development of the all-order CCSDvT (coupled-cluster singles-doubles-valence triples) method [26]. To avoid errors in coding and derivations, this complex task was aided by specially developed symbolic tools.

In 2009, the comparison of the CCSDvT and experimental results together with semi-empirical fitting tests indicated the attained accuracy at the 0.1-0.3% level [26]. Combination of the experimental value for $E1_{pv}$ with the computed structure factor $k$ led to $Q_W(^{133}Cs) = -73.16(29)_{expt}(20)_{theor}$. This was in agreement with the standard model prediction.

As is typical in a field like this that is advancing the boundaries of what is possible, the importance of various small terms and approximations in the precise calculations of *k* continue to be debated, tested, and improved upon. The authors of [20] reevaluated some sub-leading atomic structure contributions to $E_{PV}$ (contributions of the core and highly excited states to the sum (1) over intermediate states) and argued the theoretical uncertainty should be 0.5%. Although the issue is not yet settled (see critique in [12]), this work raised an important idea, whether the summation over intermediate states in the sum (1) can be avoided altogether in the coupled-cluster approach. AD is developing a new mixed-parity CCSDvT method in pursuit of this goal. Such improvements in theoretical approaches, combined with the ever-increasing power of computation, is anticipated to bring further improvements in the atomic-structure analysis. The ultimate limit in determination of the structure factor is the uncertainty in the nuclear neutron distribution [27], which introduces an error in $k$ at the level of 0.05%.

Although there has been little progress in atomic PV experiments for the past two decades, such improvements in the atomic theory is expected to stimulate further experimental work [28,29].

**Conclusion**
Atomic parity violation plays a unique, and at the same time complementary, role to collider experiments. For $^{133}$Cs atom the relevant momentum transfer is just 30 MeV, but the high accuracy probes minute contributions of the sea of virtual (including so-far undiscovered) particles at a much higher, >TeV, scale. The Cs PV results, in combination with the results of collider experiments, confirmed the energy-dependence (or ``running'') of the electroweak force over an energy range spanning four orders of magnitude, from ~ 10 MeV to ~ 100 GeV. Additionally, these result placed constraints on a variety of new physics scenarios. In particular, they increased the lower limit on the masses of extra Z-bosons predicted by models of grand unification. For some years atomic PV provided the best constraints on a number of types of extra Z-bosons. Only recently has the Large Hadron Collider improved on those constraints.

**Acknowledgment** This work was supported in part by the U.S. National Science Foundation.